# The Natural Capital Indicator Framework (NCIF): A framework of indicators for national natural capital reporting

**Authors:** A. Fairbrass[1,2]*, G. Mace[2], P. Ekins[1], B. Milligan[1,3]

**Affiliations:**

[1]Institute for Sustainable Resources, University College London, London, UK

[2]Centre for Biodiversity and Environment Research, University College London, London, UK

[3]Faculty of Law, University of New South Wales, Sydney, Australia

*Correspondence to: alison.fairbrass.10@ucl.ac.uk

**Highlights**

- A framework of natural capital indicators (NCIF) is presented.
- It includes indicators of natural capital, flows, human inputs and outputs.
- Example indicators are presented to populate the framework.
- The NCIF gives governments a structure of indicators for natural capital reporting.
- The framework is illustrated through application to national water resources.

**Summary**

It is now widely recognised that components of the environment play the role of economic assets, termed natural capital, that are a foundation of social and economic development. National governments monitor the state and trends of natural capital through a range of activities including natural capital accounting, national ecosystem assessments, ecosystem service valuation, and economic and environmental analyses. Indicators play an integral role in these activities as they facilitate the reporting of complex natural capital information. One factor that hinders the success of these activities and their comparability across countries is the absence of a coherent framework of indicators concerning natural capital (and its benefits) that can aid decision-making. Here we present an integrated Natural Capital Indicator Framework (**NCIF**) alongside example indicators, which provides an illustrative structure for countries to select and organise indicators to assess their use of and dependence on natural capital. The NCIF sits within a wider context of indicators related to natural, human, social and manufactured capital, and associated flows of benefits. The framework provides decision-makers with a structured approach to selecting natural capital indicators with which to make decisions about economic development that take into account national natural capital and associated flows of benefits.

*Keywords:* Common International Classification of Ecosystem Services (CICES), ecosystem services, environmental assets, natural capital accounts, sustainable development, System of Environmental Economic Accounting (SEEA)

**Graphical Abstract**

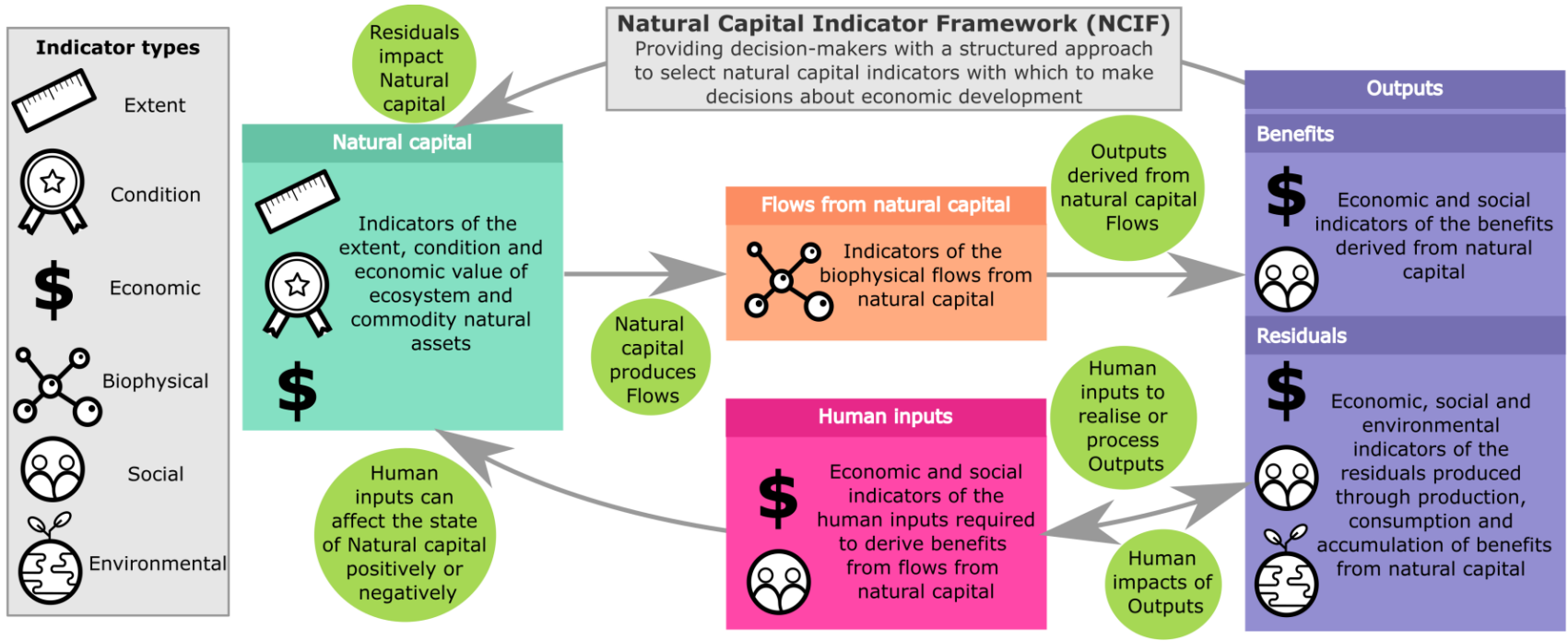

**Introduction**

Natural capital is the stock of renewable and non-renewable natural resources on earth (e.g., plants, animals, air, water, soils, and minerals) that combine to yield a flow of benefits or "services" to people [1]. Some distinctive characteristics of natural capital are that some components renew and replenish themselves, given appropriate management, and that some components are not substitutable by using other forms of capital (human, manufactured or social). Benefits from natural capital accrue ultimately from complex ecological and evolutionary processes operating across small to large spatial scales.

The importance of natural capital to development and its sustainability is recognised in the 17 Sustainable Development Goals (SDGs) [2] and 169 associated Targets, which countries have committed to achieve by 2030. The SDG Target 17.19 calls for the development of "measurements of progress on sustainable development that complement Gross Domestic Product (GDP)". This recognizes that, although GDP is the most popular and politically influential headline measure of economic progress, it gives only a partial picture of the economic status of a country or other political unit [3, 4]. For example, despite robust GDP growth since 1980, Canada's marketed natural assets (minerals, fossil fuels, timber, agricultural land and built-up land) have declined by 17% from 1980 to 2015 as a result of depletion of many of Canada's natural resources [5]. By measuring natural capital alongside GDP it will be possible to show where natural capital is being depleted and give a more comprehensive picture of a country's wealth profile.

There is an ongoing effort to develop structured concepts and accounting for relationships between the environment and the economy. Some of this effort is organised in terms of natural capital. In the public sector, the United Nations (UN) System of Environmental-Economic Accounting (SEEA) Central Framework (SEEA CF) [6], and its related components of SEEA Water, SEEA Energy, and SEEA Agriculture, Forestry and Fisheries, provide a robust environmental accounting structure which integrates with national accounting systems via the System of National Accounts (SNA). This integration enables assessment of interrelationships between the economy and the environment, including the stocks and changes in stocks of certain commodity natural capital assets, and the associated flows of goods and services. The System of Environmental-Economic Accounting – Experimental Ecosystem Accounting (SEEA EEA) [7] provides a similar structure for ecosystems and ecosystem service accounting and provides estimates of the monetary value associated with the ecosystem services that flow from their ecosystem assets. The SEEA CF is the international standard for environmental-economic accounting and has been compiled and/or published by over 80 national governments [8] while the SEEA EEA ecosystem accounts have been published in 24 countries [9] to date. In this article, we use the term 'natural capital accounting' as an umbrella terms covering efforts to use an accounting framework in a systematic way to report on stocks and flows of natural capital.

In addition to natural capital accounting, a number of scientific assessments and initiatives have generated large volumes of biophysical data that seek to illuminate the interrelationships between the environment and the economy, and that often seek to quantify the monetary value and wider economic importance of natural capital. At the country scale, national ecosystem assessments have been conducted in a number of countries including the United Kingdom [10], Portugal [11], Spain [12], and China [13] all of which have influenced the development of national policies on natural capital [14]. It is not surprising that most such assessments have been in Europe where there is a history of detailed reporting and recording on the state of the natural environment [15]. Outside Europe, many countries face significant data challenges to implementing natural capital accounting and national ecosystem assessments. The World Bank's Wealth Accounting and the Valuation of Ecosystem Services partnership (WAVES) is working with a number of countries in Africa (Botswana, Madagascar, Rwanda, Zambia), Asia (Indonesia, Philippines) and Latin America (Colombia, Costa Rica, Guatemala) to build capacity and see how such accounting can support sustainable development. The UN SEEA programme organises training and workshops in Africa [16]. In addition The Economics of Ecosystems and Biodiversity initiative (TEEB) supports countries in the valuation of natural capital [17].

A number of tools have been developed in order to conduct integrated economic and environmental analysis [18, 19]. Two of the more commonly used tools are InVest [20] and Co$ting Nature [21]. The InVest tool uses spatial data and production functions to estimate how changes in an ecosystem's structure and function are likely to affect the flows and values of ecosystem services. Different scenarios can be used to investigate the impact of different policy options, and the impacts of different scenarios are compared to inform decision-making. The

Co$ting Nature tool uses spatial datasets from remote sensing and other global sources to model biophysical and socioeconomic processes, to calculate a baseline for ecosystem services anywhere globally. Similar to the InVest tool it allows a series of interventions or scenarios of change to be modelled in order to assess their impact on ecosystem service provision. Using rapidly growing biophysical and economic datasets these tools aim to inform decision-making on natural capital and ecosystem services.

Indicators are an integral element of any system for quantifying the economy or the environment. They generally simplify in order to provide useful information about complex phenomena that can be shared among different users or different contexts. There are ongoing efforts to develop natural capital indicators, typically within broader indicator frameworks of sustainability (e.g. SDGs [2]), national wealth (e.g. World Bank Changing Wealth of Nations [22]) and green growth (e.g. Organisation of Economic Co-operation and Development (OECD) Green Growth Indicators [23]). The SEEA does not currently specify indicators for use with natural capital accounting, but the accounts do have potential to produce information to support a number of indicators from existing international indicator initiatives [24]. The OECD's Green Growth Indicators framework [23] and the Natural Capital section of the World Bank's Changing Wealth of Nations framework [22] both focus on natural capital assets, the World Bank having a stronger focus on natural resource use and the OECD having a more holistic framework that includes biodiversity. Neither framework includes indicators of ecosystems. Both are also limited in terms of how they capture the marine environment. The Natural Capital Index (NCI) currently under construction by the World Bank and the Natural Capital Project [25] takes a different approach by seeking to construct a 'production possibility frontier' from a country's natural capital, incorporating ecosystem services, measured in monetary terms, human health impacts and a biodiversity measure. The NCI would therefore permit comparisons between countries on the basis of their efficiency in making use of their natural capital endowments. A different approach again to indicators of natural capital is taken by the Intergovernmental Platform on Biodiversity and Ecosystem Services (IPBES) with their framework built around the concept of 'nature's contributions to people' (NCP) [26]; a concept that focuses on flows and benefits (and sometimes disbenefits) provided to people by nature. Notwithstanding this focus, the indicators populating the IPBES indicator framework are predominantly focused on assets, such as land cover extent and marine stocks. The Stockholm Resilience Centre has developed an indicator framework based on the SDGs and the concept of planetary boundaries [27] which is populated by some rather specific indicators that only partially cover their area of concern e.g. the suggested indicator for SDG 14 Life below water is 'Acidity of ocean surface water (pH)'. Although all of these indicator frameworks capture some components of natural capital, they tend to be limited in scope according to the context in which they have been developed. Notwithstanding all this activity, there is currently no standard set of natural capital indicators that could inform decision-making and support global efforts towards sustainable development.

Here we present a framework alongside example indicators to provide national governments with an illustrative structure to select indicators for reporting on natural capital. The Natural Capital Indicator Framework (NCIF) can incorporate the full range of a country's natural capital assets, the biophysical flows from those capital assets, the human inputs which may have co-produced these biophysical flows, the benefits deriving from those flows, and the physical residuals from them. The framework enables organization of a very large number of relevant indicators, into a coherent structure that is conducive to holistic assessment of natural capital and its interrelationships with development outcomes.

The NCIF is intended to be consistent with the conceptual framework and broad asset categories from the System of Environmental Economic Accounting (SEEA), and with the categories of flows from natural capital from the Common International Classification of Ecosystem Services (CICES) [28]. The CICES is the method of classifying ecosystem services that has been adopted by the SEEA EEA, the complementary approach to SEEA CF which approaches natural capital accounting from an ecosystem, rather than "individual environmental assets", perspective. By aligning with the SEEA and the CICES, the NCIF is designed as a coherent framework of indicators that can be populated /compiled from underlying natural capital accounts.

**A conceptual framework for natural capital**

Indicators can capture the status of natural assets, such as the extent and condition of forests and water resources. They can also be used to quantify contributions of natural capital to the formal economy, such as the net value added of timber in the national accounts, and contributions to society at large, such as the percentage of the population with access to nutritious food and safe drinking water. In order to take into account the values of nature, indicators should be measuring both the stock of a nation's natural assets, and the flows of benefits that they produce. This is often described as a natural capital approach because of the focus on the assets ('the stock') and not only the flow of ecosystem services that are represented in ecosystem service assessments [29]. A conceptual framework for natural capital indicators should therefore include all the key components of the natural capital concept: stocks (assets), flows, human inputs, and outputs in the form of benefits and residuals. Our Natural Capital Indicators Framework (NCIF) (Figure 1) is comprised of four connected components:

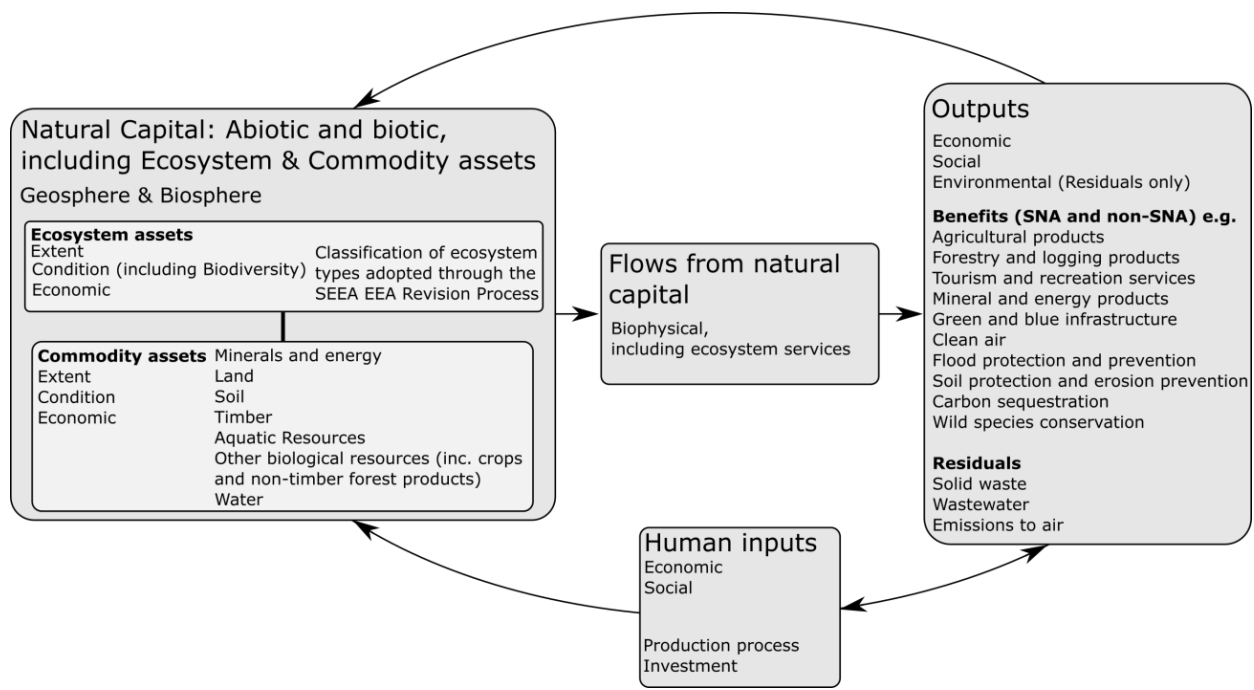

Figure 1. A conceptual framework for natural capital and the Natural Capital Indicators Framework (NCIF). SNA denotes the System of National Accounts.

**1. Natural capital:** The Earth system comprises the Geosphere and Biosphere, with the Geosphere comprising the Atmosphere, Lithosphere, Cryosphere and Hydrosphere, and the Biosphere containing all living matter that interacts with the Geosphere. Natural capital may be biotic (living systems i.e. ecosystems, animal and plant life) or abiotic (non-living matter). Within the Geosphere and Biosphere are two kinds of natural assets: ecosystem assets (including terrestrial, marine and freshwater ecosystems, with both biotic and abiotic elements, which encompass the "dynamic complex of plant, animal and micro-organism communities and their non-living environment interacting as a functional unit" [30]), and commodity assets (the environmental assets, defined in the SEEA CF, the biotic components of which are produced by ecosystem assets, and the abiotic components of which are extracted from the Geosphere).

The SEEA EEA does not define a classification of ecosystem types and this is a focus of the SEEA EEA Revision Process [31]. We will align the Ecosystem assets component of the NCIF with the ecosystem typology that will eventually be adopted by the SEEA EEA. It must be noted that a comprehensive global-scale classification of ecosystems will be complicated by the biogeographical differences among countries. There is a spatial/scaling problem (ecosystems can be overlapping at any scale) and a conceptual problem (ecosystems in different places may be functionally similar even if they are structurally quite different). It is more likely that ecosystem classification systems can be developed at the scale of countries and regions. The IUCN Red List of Ecosystems provides a methodology for classifying ecosystem types [32].

Extent, condition and economic indicators are prescribed for natural capital. Extent captures the area or quantity of each asset, condition captures the status of each asset which depends on the ecosystem service or services of interest (e.g. a good condition pasture for production may be poor for water quality), and economic captures the monetary value of the asset.

Accounting for biodiversity is important for several reasons that do not map neatly onto the natural capital framework [33]. Following the SEEA EEA, biodiversity is accounted for as part of the assessment ecosystem asset condition. Ecosystem condition metrics could include indicators of resilience and biodiversity is often a predictor of resilience [34]. While there are separate thematic accounts for species in the SEEA EEA, for simplicity these are not included in the NCIF. How to account for biodiversity is a focus of the SEEA EEA Revision Process [35] that needs to be worked out and further developed. How this evolves may affect the NCIF in the future.

Defining asset condition is important for both market and non-market ecosystem benefits and for biodiversity conservation. If the ecosystem assets are in worsening condition then the societal indicators (e.g. recreation, health, climate change resilience) and conservation benefits (fewer threatened and declining species) will show declines over time, even though other economic and social indicators might be improving. The changes in country accounts over time and the comparisons between countries should show these patterns.

**2. Flows from natural capital:** Flows include the widely understood concept of ecosystem services and our classification follows CICES [28], as already stated. Our rationale for using the flow terminology is that some users include the benefits that people receive within the definition of ecosystem services, while we are treating them as a different category in the framework because the benefits vary according to the context and user, while flows vary with assets and asset management. Also, we deliberately emphasise the distinction between assets (stocks) and flows (services). The arrow to the Outputs component from the Flows from natural capital component of the NCIF indicates that Outputs are derived from natural capital Flows.

In CICES ecosystem services are defined as the contributions that ecosystems make to human well-being that depend on either biotic or abiotic parts of ecosystems, and are distinct from the goods and benefits that people subsequently derive from them, which aligns with the NCIF. CICES is structured as a multilevel taxonomy of ecosystem services with three broad categories defined at the top level of this taxonomy, with each of these categories divided into biotic and abiotic categories: 1. Provisioning (biotic and abiotic), 2. Regulation and Maintenance (biotic and abiotic), and 3. Cultural (biotic and abiotic). This upper level of the CICES classification system can be used as a broad initial checklist suitable for different contexts [36] and supplemented with the subsequent levels of the taxonomy when more detail on particular ecosystem services is desired, making it possible for countries to adapt the framework to their specific context. We use CICES for our categorization of the flows from natural capital rather than the IPBES NCP paradigm [37] or the Final Ecosystem Goods and Services Classification System from the United States Environment Protection Agency [38] because the CICES is already acknowledged in the SEEA EEA with which we align to improve potential policy impact of the NCIF.

Not all categories of flows are relevant to every ecosystem asset. Moreover, the flows are expressed in biophysical indicators, to reflect the physical quantities. Some flows are produced by more than one asset, and some assets produce or contribute to more than one type of flow. The flows only become benefits when they acquire value for people, when they can often be expressed in monetary terms but disaggregating the contributions from different assets may not be possible. Overall, the complexity of the asset-flow-benefit causal stream, together with the difficulties in giving monetary values to non-market ecosystem goods and services, greatly increases the difficulties in valuing ecosystem assets in terms of Net Present Value (NPV).

There is also the important question of whether there are thresholds in the levels of natural capital, sometimes called 'critical natural capital' [39], below which there is a dramatic decline of, or complete cessation in, the flow of services and benefits from that capital. It would be conceptually possible to include such thresholds in the NCIF, but determining them in practice is far from straightforward [40].

**3. Human inputs:** Inputs from human activities (e.g. labour, investment, and manufactured capital) will almost always be needed alongside the natural assets in order to produce the flows from natural capital which are then experienced as benefits by people [7]. The human inputs are expressed through economic and social indicators. Economic indicators focus on the costs associated with the human inputs required to connect natural assets with benefits, while the principal social indicator associated with these human inputs is employment. The arrow from the Human Inputs component to the Natural Capital component of the NCIF indicates that human activities can impact the state of natural capital positively or negatively. The reciprocal arrow between the Human Inputs component and the Outputs component of the NCIF indicates that human inputs may be required to realise or further process the outputs from natural capital, while the Outputs have effects on humans, positive in the case of Benefits and usually negative in the case of Residuals. Residuals can also have a (normally negative) impact on Natural Capital itself, as shown by the arrow across the top of Figure 1.

**4. Outputs:** Outputs are organised into two broad categories:

1. Benefits derived from natural capital. In the context of natural capital accounting, benefits comprise: a) The value added to human welfare by the flows from natural capital (e.g., food, water, clothing, shelter, and recreation), with human inputs as required. These are referred to as System of National Accounts (SNA) benefits, since the measurement boundary is defined by the production boundary used to measure Gross Domestic Product (GDP) in the SNA. This includes goods produced from natural capital by households for their own consumption; b) The benefits that accrue to individuals that are simply flows from natural capital (e.g., clean air, flood protection from mangrove forests or coral reefs) and not produced with human inputs. These benefits are referred to as non-SNA benefits, reflecting the fact that the receipt of these benefits by individuals is not the result of an economic production process defined within the SNA. These two types of benefits may be distinguished by the fact that, in general, SNA benefits have the potential to be bought and sold on markets whereas non-SNA benefits do not [7], although limited markets may sometimes be created for non-SNA benefits through such schemes as Payments for Ecosystem Services (PES) [41]. It is important to recognise the difference between benefits and the bio-physical flows from which they are derived. The flows are bio-physical facts resulting from natural capital (e.g. flowing streams, reproduction of fish), but they only become benefits when they deliver value to people, where this value is often expressed in monetary terms. Thus, all fish stocks produce flows of fish. But only those flows of fish which give value to people are classed as benefits (while recognising that the fish may be delivering biodiversity and other ecosystem benefits and not just benefits from consumption). Benefits are assessed using economic and social indicators. Economic indicators focus on the contribution of benefits to the economy, such as the value added to the National Accounts, value associated with avoided health costs and value of mitigated damages from natural disasters. Social indicators focus on the social impacts of benefits, such as access to clean water.

2. Residuals. Residuals comprise the flows of solid, liquid and gaseous materials, and energy that are discarded, discharged or emitted by establishments and households through processes of production, consumption or accumulation [6].

Residuals are assessed using economic, environmental and social indicators. Economic indicators focus on the costs of processing residuals or the damages caused by them, environmental indicators focus on volumes of residuals, and social indicators focus on the social impacts of residuals, such as the percentage of a population exposed to dangerous levels of air pollution.

**Context of the Natural Capital Indicator Framework (NCIF)**

The NCIF sits within a wider context of indicators related to natural, human, social and manufactured capital, and associated flows of benefits. Figure 2 illustrates a four-capital model of wealth creation which was first put forward in [42] and elaborated further in [43]. The greyed boxes in Figure 2 indicate the components related to natural capital, and include the categories that comprise the Natural Capital Indicator Framework presented in this paper. Figure 2 portrays four kinds of capital stock: ecological (or natural) capital, human capital, social and organisational capital, and manufactured capital. Each of these stocks produces a flow of 'services' from the environment ($E$), from human capital ($L$), from social/organisational capital ($S$), and from physical capital ($K$), services which serve as inputs into the productive process, along with 'intermediate inputs' ($M$), which are previous outputs from the economy and are used as inputs in a subsequent process. Other types of capital have been put forward, principally among them financial capital. However, financial capital, and the financial system through which it acts, may better be seen as a type of social capital, a conventional way of allocating and representing the power to mobilise the other four kinds of capital which have the real inherent power to deliver benefits.

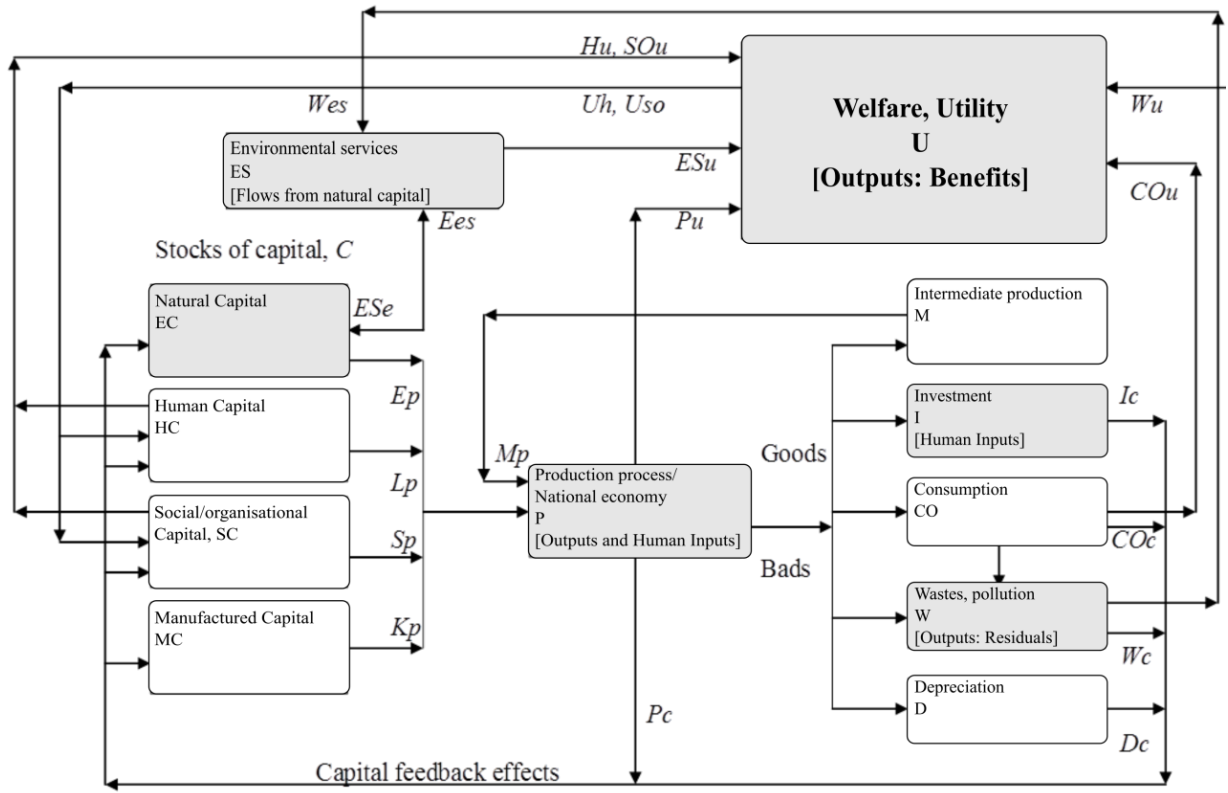

Figure 2. Four-Capital Model of Wealth Creation through a Process of Production adapted from [42]. Grey boxes highlight the components that are reflected in the Natural Capital Indicator Framework developed in this paper (Natural Capital, Flows from Natural Capital, Human Inputs and Outputs including Benefits and Residuals). In the flow descriptors, the upper case letters denote the source of the flow, lower case letters denote the destination. Those relating to the various capital stocks have the C omitted for simplicity.

The Natural Capital Indicators Framework (Figure 1) is closely aligned with existing frameworks of natural capital [29] and national natural capital accounting [44]. It also has several noteworthy points of contrast with recent literature, in particular with the natural capital asset classification recently presented by Leach et al. (2019)[45]. The major points of contrast are:

- Unlike Leach et al., Figure 1 makes no clear distinction between biotic and abiotic assets. The classification here is based on the definable flows of services and benefits into the economy – this corresponds to the definition of capital. It is also necessary to have

- interacting biotic and abiotic components in asset classes in order that they deliver their functional roles, for example natural capital assets (e.g. soil, ecosystems) have mixed biotic and abiotic elements.
- Again unlike Leach et al., the NCIF in Figure 1 treats biodiversity as a characteristic of all ecosystems, which are in the top level of natural capital, rather than as a distinct asset. Biodiversity is a key indicator of ecosystem asset condition in the NCIF. Clearly the flows from natural capital, and the benefits they result in, are dependent on the characteristics of ecosystems, including biodiversity, although the relationships and roles of the different characteristics in producing the flows are complex.
- Finally, Figure 1 identifies the flows from and benefits of natural capital as core parts of the natural capital indicator framework. In Leach et al. they appear as isolated case study examples. Yet it is the flows and benefits that actually distinguish natural capital from environmental components of no economic interest. This is important because it is the trend in the flows and benefits from natural capital that are relevant to questions as to whether the natural capital is being used sustainably or not, if necessary reflecting lags and thresholds between asset condition, flows and benefits

There remains uncertainty about how biodiversity should be included in natural capital accounts [46]. This is mostly because biodiversity is such a broad term and is often used vaguely for assets, services and benefits. However, if biodiversity components are clarified then it is clearly either an asset or a benefit (and sometimes a service itself) [33, 47]. In our framework, we include biodiversity as a measure of ecosystem asset condition, following the SEEA EEA. The conservation of wild species is also included as a benefit. In order to achieve this benefit we need to see both the diversity (number of species) and abundance of wild species at least being maintained and sometimes increasing. Therefore indicators of species abundance (Living Planet Index [48]) and diversity (Red List Index [49]) are included as flows within the framework.

**Examples of natural capital indicators**

In Table 1 we present examples of indicators that can be used to populate the NCIF, to provide guidance on the appropriate indicators to select when applying the NCIF. The purpose of the indicators is: 1. To provide public policy-makers with summary information about the state, condition and value of natural capital assets, flows from natural capital, human inputs, and outputs including benefits from these flows and residuals, 2. To provide a set of indicators for natural capital that can operate as a front-end for a system of natural capital accounting such as SEEA, and 3. To assess if development is occurring sustainably.

Table 1. Descriptions and examples of the types of indicators to populate each component of the Natural Capital Indicators Framework. As already noted, in CICES ecosystem services are classified as provisioning, regulation and maintenance, or cultural, and as either biotic or abiotic. In the examples below these distinctions have been omitted for simplicity.

| Component | Indicator Type | Indicator Description | Examples |
|---|---|---|---|
| *Commodity assets* | Extent | The quantity of the asset type measured by volume or area | Reserves of mineral and energy resources; Area of timber resources |
| | Condition | The condition of the asset type | Energy return on energy investment (EROEI) (mJ/t); Wood quality indicator |
| | Economic | The net present value of the asset type | Net present value of mineral and energy reserves ($) |
| *Ecosystem assets* | Extent | The quantity of the asset type measured by volume or area | Area of forest ecosystem assets (ha) |
| | Condition | The condition of the ecosystem asset measured by an index of biodiversity | Biodiversity Intactness Index for forest ecosystem assets |
| | Economic | The net present value of the asset type | Net present value of terrestrial ecosystem assets |
| *Flows* | Biophy-sical | The biophysical flows of ecosystem services measured by volume, area or an index of biodiversity | Volume of mineral and energy resources extracted (tonnes); Change in area of different types of land use; Living Planet Index; Red List Index |
| *Human inputs* | Economic | The financial cost of deriving benefits from natural assets via ecosystem services measured by costs of cultivation, management and/or extraction of natural resources | Cost of harvesting timber ($); Expenditure in managing soil erosion ($); Cost of managing terrestrial ecosystem assets ($) |

|  | Social | The human capital required to derive benefits from natural assets via ecosystem services measured by employment in cultivation, management and/or extraction of natural resources | Percentage of population employed in the timber industry (%) |
|---|---|---|---|
| *Output: Benefits* | Economic | The financial benefits derived from natural capital via flows of ecosystem services. Economic benefits are measured by the gross value added in the National Accounts, land rents, avoided costs from natural disasters, or value of markets in natural resources | Gross value added in the National Accounts associated with mineral and energy resources ($); Land rents ($); Costs of water related damage (floods, coastal damage) ($); Value of jewellery market ($) |
|  | Social | The social benefits derived from natural capital via flows of ecosystem services. Social benefits are measured by access to natural resources, impacts of natural disasters, exposure to pollution, or engagement with natural capital | Percentage of population with access to safe water (%); Percentage of population affected by water-related events (%); Percentage of population who are members of biodiversity conservation organisations (%) |
| *Output: Residuals* | Economic | The economic costs of waste and pollution produced through the process of deriving benefits from natural capital. Costs are measured by expenditure on waste disposal and pollution treatment and damages | Cost of solid waste treatment ($); Damages from stratospheric ozone depletion ($) |
|  | Social | The social impacts of residuals produced through the process of deriving benefits from natural capital. Impacts are measured by employment in related industries and health impacts of residuals | Percentage of population employed in the wastewater industry (%); Percentage of population exposed to water pollution (%) |
|  | Environ-mental | The quantity and environmental impact of residuals produced through the process of deriving benefits from natural capital. Impacts are measured by amount of residuals produced, managed and emitted into the environment. | Volume of waste managed by management type (tonnes); GHG emissions (tonnes) |

A fully populated NCIF would contain a large number of indicators, comprehensively describing different natural capital assets, flows from natural capital, human inputs into natural capital, and outputs from natural capital including benefits from natural capital and residuals that may affect natural capital or the benefits derived from it. These indicators may be compared with indicators from scientific literature on indicators and other international indicator initiatives such as those associated with the Sustainable Development Goals and Aichi Targets. A full set of suggested indicators and some international comparators are given in [50].

**Making the NCIF Operational**

The NCIF presented here provides national governments with an illustrative structure to select indicators for reporting on natural capital. The framework enables organization of a very large number of relevant indicators, into a coherent structure that is conducive to holistic assessment of natural capital and its interrelationships with development outcomes. We suggest that users carefully select a set of indicators, relevant to their context, from each component of the framework, to ensure that a comprehensive set of indicators is selected. For example, when reporting on national water resources a government should select indicators relevant to water assets from each component of the NCIF (Figure 3). This illustrates the policy relevance of the NCIF to national governments.

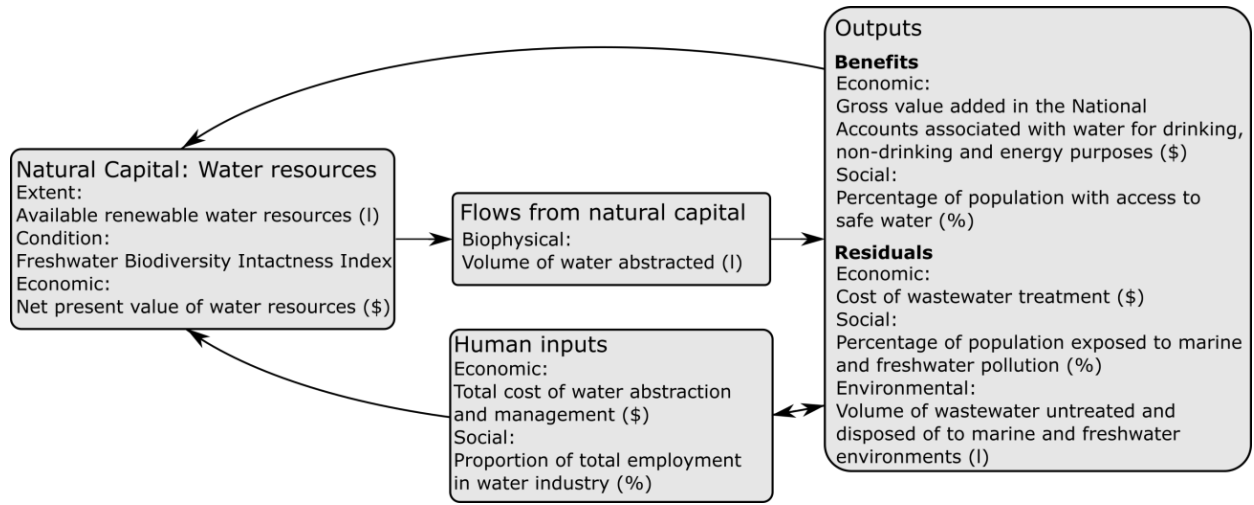

Figure 3. Illustrative example of indicators to monitor national water resources. Potential indicators are proposed for each component of the Natural Capital Indicators Framework.

We hope that the framework will show how important defining asset condition is for both market and non-market ecosystem benefits and for biodiversity conservation. If the ecosystem assets are in poor condition then the societal indicators (e.g. recreation, climate change resilience) and conservation benefits (fewer threatened and declining species) will show declines over time while other indicators might be increasing. The changes in country accounts over time and the comparisons between countries should show these patterns.

For the NCIF to become widely used an international process will need to build in this work by refining and formalising a coherent and flexible framework of natural capital indicators that are specifically tailored to the practical requirements of policy decision-making about sustainable development. Some of the suggested steps in that process are to:

- *Develop guiding principles for using the framework:* A set of guiding principles should be developed to accompany the NCIF to support potential users in its application. This could include how to select the most appropriate indicators for the framework for specific contexts, how to apply the NCIF for different user types e.g. public vs. private use cases, and how to use the NCIF to incorporate natural capital into existing indicator frameworks. These principles could be developed during the testing of the potential applications and use cases of the NCIF. In due course they could be formalized into a set of standards to allow consistency and comparability among countries in natural capital measurement, leading to an overall understanding of the state of natural capital.
- *Define criteria for selecting indicators:* Criteria should be defined for selecting indicators for the NCIF, in order to: 1. Provide more robustness to the choice of indicators in the NCIF, and 2. Guide users in the development of new indicators when existing available indicators are not fit-for purpose, including issues such as data availability, thresholds, critical values and uncertainty ranges. Unfortunately the majority of existing indicators are far from ideal for monitoring natural capital assets, flows from natural capital, human inputs into natural capital, and outputs from natural capital. In practice, practitioners compiling natural capital accounts will have to choose between using those indicators that are available but not necessarily fit for purpose, or setting out to develop a fit-for-purpose indicator.

- *Identify, develop and organize natural capital sub-indicators:* Our review of indicators presented in [50] highlights several potential coverage gaps in the current availability of natural capital indicators, which would need to be addressed to maximise the coverage and practical utility of the framework. Further work is needed to develop indicators to cover all significant ecosystems and other natural resources [51]. For example, there is a need for collaboration to identify, develop and organize specific indicators for: biodiversity as an indicator of asset condition; regulation and maintenance services generally; and the extent, condition and associated flows for marine assets generally.

- *Develop understanding of relationships between indicators across the framework:* The relationship between indicators across the multiple components of the framework could be used to infer information about the state of stocks and flows. For example, if human input indicators increase while benefit indicators remain constant, this may signal that the stock is degraded and requires increasing human effort to extract the same amount of flows and benefits. This information would be useful to inform decisions and monitor the impacts of policies.

- *Develop understanding of how trade-offs of different ecosystem services are captured by the NCIF:* There are likely to be trade-offs between different ecosystem services, and it should be understood whether and how this is captured by the NCIF. For example, timber extraction from a forest ecosystem may increase the biotic provisioning flow of timber resources while reducing multiple flows from forest ecosystems, including regulation and maintenance and cultural flows.

- *Identify practical use cases for the framework and indicators:* Natural capital indicators can be used at different levels in order to facilitate decision-making:
    - At the inter-governmental level, elements of the NCIF could be adapted as appropriate to embed a natural capital perspective within broader indicator frameworks of sustainable development and green growth, for example those maintained by the OECD [52], World Bank [53], and other multilateral institutions.
    - At a national level, elements of the NCIF could be embedded as appropriate within national indicator frameworks for sustainable development; progress reporting for the

> SDGs and other international commitments; and within economic performance assessment generally as a contextualising complement to GDP.

**Conclusion**

Natural capital is an economic asset that underlies social and economic development. International commitments including the Agenda 2030 emphasise the need for national governments to value and account for natural capital in decision-making to avoid economic development that is dependent on unsustainable depletion of natural resources. Despite a range of initiatives and tools to do this, such as natural capital accounting, there currently exists no comprehensive approach to natural capital indicators for national natural capital reporting. Here we present a Natural Capital Indicator Framework (NCIF) based on the concept of capital organized around four linked components: stocks (assets), flows, human inputs, and outputs in the form of benefits and residuals. Alongside the NCIF we present examples for natural capital indicators in order to provide guidance on the appropriate indicators to select when applying the NCIF. This framework provides a structured approach for governments to select a holistic suite of natural capital indicators for national reporting that are appropriate to their context. Future work is required to develop indicators for biodiversity as a condition of natural assets, regulation and maintenance ecosystem services, and marine assets. Inter-relationships between indicators across the NCIF may highlight issues such as efficiency and resource depletion and this needs to be investigated. Guidance on applying natural capital indicators will be required to allow consistency and comparability among countries and an overall understanding of the state of natural capital, and the NCIF needs to be pilot tested to understand in what governance contexts it is useful. To avoid ongoing economic development that relies on unsustainable natural capital depletion national governments need to move beyond traditional indicators of economic development to include reporting of indicators on the full suite of capitals, including natural capital. The work presented here provides national governments with a structured approach to selecting indicators for this purpose.

**Acknowledgements**

This research was conducted in partnership with the Green Growth Knowledge Platform (GGKP) Research Programme with generous support from the MAVA foundation. We thank the members of the GGKP Natural Capital Working Group who provided comments on earlier versions of this work, and the attendees of a workshop held at the United Nations in Geneva in July 2019 who provided feedback on this research.